\documentclass[fleqn,twoside]{article}
\usepackage[headings]{ll}
\usepackage{axodraw}
\usepackage{colordvi}
\newcommand{\nl}{\nonumber\\}

\newcommand{\lpar}{\left(}                            % bracketing
\newcommand{\rpar}{\right)}

\newcommand{\bq}{\begin{equation}}                    % equationing
\newcommand{\eq}{\end{equation}}
\newcommand{\bqa}{\arraycolsep 0.14em\begin{eqnarray}}
\newcommand{\eqa}{\end{eqnarray}}
\newcommand{\ba}[1]{\begin{array}{#1}}
\newcommand{\ea}{\end{array}}
\newcommand{\ben}{\begin{enumerate}}
\newcommand{\een}{\end{enumerate}}
\newcommand{\bei}{\begin{itemize}}
\newcommand{\eei}{\end{itemize}}
\newcommand{\eqn}[1]{Eq.(\ref{#1})}

\def\Re{\mathop{\operator@font Re}\nolimits}
\def\Im{\mathop{\operator@font Im}\nolimits}

\newcommand{\mws}{M^2_{_W}}

\newcommand{\mhs}{M^2_{_H}}

\newcommand{\mts}{m^2_t}

\newcommand{\spro}[2]{{#1}\cdot{#2}}
\newcommand{\intmomi}[2]{\int\,d^{#1}#2}

\newcommand{\Reb}{{\rm{Re}}}
\newcommand{\Imb}{{\rm{Im}}}
\newcommand{\upar}[1]{u}

\newcommand{\ssA}{{\scriptscriptstyle{A}}}

\newcommand{\ssD}{{\scriptscriptstyle{D}}}

\newcommand{\ssN}{{\scriptscriptstyle{N}}}

\newcommand{\ssT}{{\scriptscriptstyle{T}}}

\newcommand{\bqas}{\begin{eqnarray*}}
\newcommand{\eqas}{\end{eqnarray*}}
\def\app#1#2 {{\it Acta. Phys. Pol.} {\bf#1},#2}
\def\cpc#1#2 {{\it Computer Phys. Comm.} {\bf#1},#2}
\def\np#1#2 {{\it Nucl. Phys.} {\bf#1},#2}
\def\pl#1#2 {{\it Phys. Lett.} {\bf#1},#2}
\def\prep#1#2 {{\it Phys. Rep.} {\bf#1},#2}
\def\prev#1#2 {{\it Phys. Rev.} {\bf#1},#2}
\def\prl#1#2 {{\it Phys. Rev. Lett.} {\bf#1},#2}
\def\zp#1#2 {{\it Zeit. Phys.} {\bf#1},#2}
\def\sptp#1#2 {{\it Suppl. Prog. Theor. Phys.} {\bf#1},#2}
\def\mpl#1#2 {{\it Modern Phys. Lett.} {\bf#1},#2}
\def\jetp#1#2 {{\it Sov. Phys. JETP} {\bf#1},#2}
\def\fpj#1#2 {{\it Fortschr. Phys.} {\bf#1},#2}
\def\afp#1#2 {{\it Acta.Phys. Polon.} {\bf#1},#2}
\def\err#1#2 {{\it Erratum} {\bf#1},#2}
\def\ijmp#1#2 {{\it Int. J. Mod. Phys} {\bf#1},#2}
\def\nc#1#2 {{\it Nuovo Cimento} {\bf#1},#2}
\def\ap#1#2 {{\it Ann. Phys.} {\bf#1},#2}
\def\cmp#1#2 {{\it Comm. Math. Phys.} {\bf#1},#2}
\def\el#1#2 {{\it Europhys. Lett.} {\bf#1},#2}
\def\hpa#1#2 {{\it Helv. Phys. Acta} {\bf#1},#2}
\def\yf#1#2 {{\it Yad. Fiz.} {\bf#1},#2}
\def\nim#1#2 {{\it Nucl. Instrum. Meth.} {\bf#1},#2}
\def\spz#1#2 {{\it Sov. Pisma Zhetf} {\bf#1},#2}
\def\jetpl#1#2 {{\it JETP Lett.} {\bf#1},#2}
\def\sjnp#1#2 {{\it Sov. J. Nucl. Phys.} {\bf#1},#2}
\def\ptp#1#2 {{\it Progr. Theor. Phys. (Kyoto)} {\bf#1},#2}
\def\rmp#1#2  {{\it Rev. Mod. Phys.} {\bf#1},#2}
\def\zhetf#1#2 {{\it ZhETF} {\bf#1},#2}
\def\prs#1#2 {{\it Proc. Roy. Soc.} {\bf#1},#2}
\def\phys#1#2 {{\it Physica} {\bf#1},#2}

\def\bfi{\begin{figure}}
\def\efi{\end{figure}}

\newcommand{\bmid}{\Bigr|}
\usepackage{graphicx}
\usepackage[figuresright]{rotating}

\newcommand{\AmS}{{\protect\the\textfont2
  A\kern-.1667em\lower.5ex\hbox{M}\kern-.125emS}}

\title{Anomalous Threshold as the Pivot of Feynman Amplitudes}

\author{
Stefano~Goria\address[TO]{Dipartimento di Fisica Teorica, Universit\`a 
                                di Torino, Italy
}
and 
Giampiero~Passarino\addressmark[TO]
\thanks{Work supported by MIUR under contract
2001023713$\_$006, by INFN and by the European Community's Marie-Curie Research 
Training Network under contract MRTN-CT-2006-035505
`Tools and Precision Calculations for Physics Discoveries at Colliders'.}}
\runauthor{S. Goria, G. Passarino}

\begin{document}

\begin{abstract}
Reduction techniques, Landau singularities and differential equations for Feynman 
amplitudes are briefly reviewed. 
\vspace{1pc}
\end{abstract}

\maketitle
%--
\section{Reduction techniques and factorization}
%--
A modern version of reduction of Feynman integrals~\cite{Cachazo:2008vp}
tells us that 
%--
\bqa
\sum\,{}&{}&{}\Bigl\{ \hbox{1-loop n-legs Feynman diagrams} \Bigr\} =
\nl
{}&{}&{} \sum_{\cal D}\,B_{\cal D}\,
D_0\Bigl( P^{\cal D}_1,\,\dots\,,P^{\cal D}_4\Bigr) + \cdots
\eqa
%--
where ${\cal D}$ is a partition of $\{1\dots n\}$ into $4$ non-empty sets,
$P^{\cal D}_i$ is the sum of momenta in $i \in {\cal D}$ and $D_0$ a
scalar box.
In other words, scalar one-loop integrals (up to boxes) form a basis. Thus, 
coefficients in the expansion ($B_{\cal D}$ etc.) are uniquely determined, 
although some reduction method can be more efficient than others. 
However, troublesome points where the numerical stability of the result is
at stake will always be there. What to do in these cases? We can change (adapt) 
bases, or avoid bases (expansion). 

We explain our idea via examples; first, we consider factorization of Feynman 
amplitudes, the Kershaw theorem of Ref.~\cite{Kershaw:1971rc}:
any Feynman diagram is particularly simple when evaluated around its anomalous 
threshold. The singular part of a scattering amplitude around its leading Landau 
singularity may be written as an algebraic product of the scattering
amplitudes for each vertex of the corresponding Landau graph times a 
certain explicitly determined singularity factor which depends only on the
type of singularity (triangle graph, box graph, etc.) and on the masses 
and spins of the internal particles.

Let us illustrate the consequences of factorization with one example: define a 
scalar one-loop $N\,$-leg integral in $n\,$-dimensions as
%--
\bqas
N^n_0= \lambda_n\,\int\,\frac{d^nq}{\prod_{i=0}^{N-1}\,[i]},
\quad [i] = P^2_i+m^2_i,
\eqas
%--
with $\lambda_n= \mu^{4-n}/(i\,\pi^2)$ and $P_i= q + \,\dots\,+ p_i$
($p_0= 0$). In parametric space we have
%--
\bqa
N^n_0 &=& \lpar\frac{\mu^2}{\pi}\rpar^{N-n/2}\,\Gamma \lpar N-\frac{n}{2}\rpar\,
{\cal N}^n_0,
\nl
{\cal N}^n_0 &=& \prod_{i=1}^{\ssN}\,\int_0^{x_{i-1}}\,dx_i\,V^{n/2-N}_{\ssN},
\eqa
%--
\bqas
V_{\ssN} &=& x^t\,H_{\ssN}\,x + 2\,K^t_{\ssN}\,x + L_{\ssN},
\quad
X_{\ssN} = - K^t_{\ssN}\,H^{-1}_{\ssN}. 
\eqas
%--
Standard notation for $N= 1,2\,\dots$ is $N_0= A_0, B_0\,\dots$; the superscript 
$n$ will be dropped unless strictly needed.
In order to discuss the procedure it is helpful to introduce the following
quantities: the BST factor~\cite{BST}, 
$B_{\ssN} = L_{\ssN} - K^t_{\ssN}\,H^{-1}_{\ssN}\,K_{\ssN}$,
the Gram matrix, $H_{\ssN,ij} = -\,\spro{p_i}{p_j}$
the Caley matrix~\cite{Melrose:1965kb}
%--
\[
M_{\ssN} =
\left(
\begin{array}{cc}
H_{\ssN}   & K_{\ssN} \\
K^t_{\ssN} & L_{\ssN} \\
\end{array}
\right)
\]
%--
It follows that~\cite{Ferroglia:2002mz} $B = C/G$ for any $N$, where 
$C = {\rm det} M$ and $G= {\rm det H}$. Landau singularities are seen as pinches 
(we assume that masses and invariants $\,\in R$) when we write 
$V_{\ssN} = \lpar x - X_{\ssN} \rpar^t\,H\,\lpar x - X_{\ssN} \rpar + B_{\ssN}$.
This realtion indeed shows that $B_{\ssN} = 0$ is the origin of the pinch 
on the integration contour at the point of coordinates $x= X_{\ssN}$; 
therefore, if the conditions, $B_{\ssN} = 0$ and 
$0 < X_{\ssN,\ssN-1} < \,\dots\,< X_{\ssN,1} < 1$, are satisfied we will have the 
leading singularity of the diagram (hereafter AT). 

Nowadays, the keyword in any reduction procedure is to avoid inverse powers
of Gram determinants. A common wisdom, but why? The vanishing of the Gram 
determinant is the condition for the occurrence of non-Landau singularities, 
connected with the distorsion of the integration contour to infinity; furthermore, 
for complicated diagrams (see Sect.~10 of Ref.~\cite{Bern:2008ef}), there may be 
pinching of Landau ($C= 0$) and non-Landau singularities ($G = 0$), giving 
rise to a non-Landau singularity whose position depends upon the internal 
masses~\cite{elop}.

Given the above properties, the factorization of Kershaw 
theorem~\cite{Kershaw:1971rc} follows.
The beauty of being at the anomalous threshold is that scalar products are frozen 
as a consequence of the Landau equations and the amplitude factorizes. Therefore, 
the AT looks perfect for boundary conditions, as long as it is inside the
physical region. Alternatively we may expand and match 
residues at a given AT~\cite{Cachazo:2008vp}. 

Let us consider standard reduction~\cite{Passarino:1978jh} as compared to modern 
techniques~\cite{Bern:1995mk} by taking a box diagram with $\spro{q}{p_1}$ in 
the numerator:
%--
\bq
\spro{D}{p_1} =
\sum_{i=1}^3\,D_{1i}\,\spro{p_1}{p_i} = 
-\,\sum_{i=1}^3\,D_{1i}\,H_{1i}.
\eq
%--
A careful application of the standard method gives
%--
\bqas
D_{1i} &=& -\,\frac{1}{2}\,H^{-1}_{ij}\,d_j,
\;\;
d_i = D^{(i+1)}_0 - D^{(i)}_0 - 2\,K_i\,D_0,
\eqas
%--
where $D^{(i)}_0$ is the scalar triangle obtained by removing propagator 
$i$ from the box. Therefore we obtain
%--
\bq
\spro{D}{p_1} =
\frac{1}{2}\,\sum_{i,j=1}^3\,H^{-1}_{ij}\,H_{1i}\,d_j = \frac{1}{2}\,d_1,
\eq
%--
without having to introduce $G_3$. Furthermore, the coefficient of the scalar $D_0$
in the reduction is $1/2\,(m^2_0 - m^2_1 - p^2_1)$.
At the AT of the box we must have $q^2 + m^2_0 = 0$, $( q+p_1)^2 + m^2_1 = 0$, 
etc. Therefore the coefficient of $D_0$ is fixed by
%--
\bq
2\,\spro{q}{p_1}\;\bmid_{\ssA\ssT} = m^2_0 - m^2_1 - p^2_1,
\eq
%--
which is what a careful application of standard reduction gives. Note that one 
gets the coefficient without having to require a physical singularity.
In standard reduction for a $N\,$-point function each, reducible, scalar
product in the numerator is replaced by a difference of propagators plus
a $K\,$-factor. The latter is what is predicted by factorization at the
anomalous threshold; the procedure is continued and one finds $N-1\,$point
functions with reducible and also irreducible scalar products; for the latter
inverse powers of Gram determinants remain.

It is worth noting that starting from six legs factorization must be understood 
as performed at some subLeading Landau singularity of the diagram~\cite{lands}, 
which is easily achieved by using the BST-algorithm~\cite{BST}. If the derivation 
is to hold we must further require that the leading Landau singularity point does 
not also lie on the Landau curve of its sub-graphs.
For illustration, consider a box in $n\,$-dimensions in a region where 
$B_4 \not= 0$. BST relations allow us to decompose the box in a 
$n+2\,$-dimensional box plus four $n\,$-dimensional triangle, 
$D^n \to D^{n+2}\,\oplus\,C^n$. A second application gives 
$D^{n+4}\,\oplus\,C^{n+2}\,\oplus\,C^n$. A box in $8\,$-dimensions as well as 
a triangle in $6\,$-dimensions cannot develop a singularity, threfore the 
subleading singularities of the original box are given by the leading ones of the 
four triangles obtained by shrinking one of the lines in the box to a point.
The coefficients of the decomposition can be found in~\cite{Ferroglia:2002mz}
and the argument can be generalized to arbitrary number of legs.

To summarize, at least in one {\em point} we can avoid reduction, all integrals 
are scalar; however, we need to have the AT inside the physical region 
$R_{\rm phys}$ (support of $\Delta^{\pm}\,$-propagators in $R$)
Since this is a rare event we must have a generalization of the factorization
theorem: prove that the AT, even with invariants 
$\,\not\in\,R_{\rm phys}$ implies a frozen $q$.

If a one-loop, $N\,$-legs scalar diagram is singular at $x= X_{\ssN} \in R$ 
then consider $N^{n\mu}\,p_{\mu\,l}$,  
%--
\bqas
\spro{N^n}{p_l} &=& -\sum_{i=1}^N\,{\cal N}^n(i)\,\spro{p_l}{p_i} 
\nl
\sum_{i=1}^N\,{\cal N}^n(i)\,H_{li}
&\stackrel{\sim}{\ssA\ssT}&
\sum_{i=1}^N\,{\cal N}^n(1)\,H_{li}\,X_i = -\,K_l\,{\cal N}_n.
\eqas
%--
where ${\cal N}^n(i)$ is the same as the scalar integral (${\cal N}^n(1)$) but 
with one power $x_i$ in the numerator, and $H\,X = - K$: this leads to generalized 
factorization since, at the AT, all scalar products are replaced by the solution of 
$(q + \dots + p_i)^2 + m^2_i = 0$, with $i= 0,\,\dots\,,N-1$.
%--
\section{Feynman diagrams aroud AT}
%--
In this section we consider a classification of {\em physical} ATs:
for instance, direct calculation shows that, for $N= 4$, there are $14$ branches 
in $p\,$-(real) space. In general, this classification is much  easier when we use 
the Coleman - Norton theorem~\cite{Coleman:1965xm}.
As a consequence of it, in a $2 \to 2$ process, two unstable particles in the 
initial state are needed. Other simple examples of physical AT are represented by
a) $\gamma^* (Z^*) \to {\bar b} b H$ (for a virtuality $s > 4\,m^2_t$ and 
$M^2(H) > 4\,\mws$) and b) from pentagons arising in the reduction of the
hexagon in $\gamma^* (Z^*) \to {\bar b} b {\bar\nu} \nu H$ (as suggested by 
A.~Denner).

The expansion of Feynman integrals around their AT is easy to derive analytically 
and only requires Mellin-Barnes and sector decomposition techniques as
explained in Ref.~\cite{Ferroglia:2002mz}.
Examples of leading behavior are:
for the vertex $C_0 \sim \ln\,B_3$; for the box $D_0 \sim B^{-1/2}_4$; 
for the pentagon $E_0 \sim B^{-1}_5$ and no singularity 
for the hexagon $F_0$ in $4\,$dimensions~\cite{lands}; e.g. $\;\Imb\,C_0$ has a 
logarithmic singularity, $\;\Reb\,C_0$ has a discontinuity.
Here we do not consider infrared/collinear configurations where we expect an
enhancement of the singular behavior (in the residues of IR/coll. poles).

It is worth noting the non-integrable (scalar) pentagon singularity which seems 
to require the introduction of complex masses for unstable internal 
particles~\cite{Denner:2006ic}.
For integrable singularities we always average over a Breit-Wigner of the invariant 
mass of unstable external particles.
%--
\section{Differential equations}
%--
An interesting feature of factorization at AT is the possibility
of introducing a differential equation with boundary conditions at the AT where
the amplitude is directly given in terms of scalar functions; what 
we want is an ODE for the full amplitude, with real momenta and one boundary 
condition; this requires to find the right variable.
The advantages of this procedure are given by a total absence of reduction and 
by the extedibility to higher loops.

It is well-known that non-homogeneous systems of ODE~\cite{Argeri:2007up}
are easy to obtain with IBP-techniques~\cite{IBPI} but the non-homogeneous part 
requires (a lot) of additional work; the natural alternative would be to introduce
PDE. They are notoriously much more difficult to handle even if
homogeneous (compatible) systems of nth-order PDE are easy to derive, a fact that 
has to do with the hypergeometric character of one-loop diagrams.
It is enough to use Kershaw expansion around pseudo-threshold~\cite{Kershaw:1973km}
and a generalization of Horn-Birkeland-Ore theory~\cite{ellip}. 
%--
\section{Diffeomorphisms}
%--
Let us restrict to ODE.
To achieve our goal we find it most natural to introduce special diffeomorphisms 
${\cal T}$ of the 
Feynman diagrams. Define $P_i(z)= T_{ij}(z)\,p_j$ with $\sum\,P_i = \sum\,p_i = 0$ 
and with $T_{ij}(0)= \delta_{ij}$; next we look for a
$z = z_{\ssA\ssT} \in R$ where the transformed diagram is singular.
Furthermore, ${\cal T}$ is {\em physical} if maps $D(0)$ onto a $D(z)$ which is 
singular at $z_{\ssA\ssT} \in R$ and $s_{ij} \to S_{ij}(z) \in {\rm Phys}_z$,
where $s_{ij}$ and $S_{ij}$ are invariants; no restriction on $s_{ij}$ is required.
${\cal T}$ is {\em unphysical} if maps $D(0)$ onto a $D(z)$ which is singular 
at $z_{\ssA\ssT} \in R$ but $s_{ij} \to S_{ij}(z) \not\in {\rm Phys}_z$; it 
requires restrictions on the original invariants $s_{ij}$.

A general solution of our problem is as follows: 
if $\,\exists\;\;$ a diagram ${\overline D}$, 
a transformation ${\overline T}$ such that
${\overline D}(z) = {\overline T}(z)\,{\overline D}$ with 
${\overline T}(0) = I$ and 
${\overline D}(z_{\ssA\ssT})$ singular ($z_{\ssA\ssT} \in R$)
then we map $D$ as follows:
%--
\bqa
D &\to& D\lpar z,z_{\ssA\ssT}\rpar
\nl
D\lpar z,z_{\ssA\ssT}\rpar &=& 
T_1\lpar z,z_{\ssA\ssT}\rpar\,D + T_2\lpar z,z_{\ssA\ssT}\rpar\,
{\overline D}(0)
\nl
T_1\lpar 0,z_{\ssA\ssT}\rpar &=& I, \quad 
T_2\lpar 0,z_{\ssA\ssT}\rpar = 0
\nl
T_1\lpar z_{\ssA\ssT},z_{\ssA\ssT}\rpar &=& 0, \quad 
T_2\lpar z_{\ssA\ssT},z_{\ssA\ssT}\rpar = I.
\label{GM}
\eqa
%--
It is worth mentioning that, in this way, we can write a differential equation
for the full amplitude instead of one for each master integral with different
boundary conditions. The interesting feature can be summarized as follows:
for a given {\em topology} which is candidate to satisfy Coleman - Norton 
(e.g. crossed box in $2 \to 2$) we perform the transformation in such a way
that the new invariants indeed satisfy the conditions of the theorem; for all
parent {\em topologies} (e.g. direct boxes) we use the general mapping
described in \eqn{GM}.

It is straightforward to see how our approach is related to the one of
differential equations in Mandelstam variables:
%--
\bq
\tau^{-1}_{ij}\,\frac{d}{dz} = O_{ij}\lpar \{s\}\rpar =
P_{i\mu}\,\frac{\partial s_l}{\partial P_{j\mu}}\,\frac{\partial}{\partial s_l},
\eq
%-- 
where $\tau= (dT/dz)\,T^{-1}$.
 
As an example for a four-point function we consider one of the crossed
diagrams in $gg \to \gamma \gamma$ with a massive loop. The transformation is
%--
\[
T = 
\left(
\begin{array}{cccc}
1-z & 0   & z   & 0 \\
0   & 1-z & 0   & z \\
z   & 0   & 1-z & 0 \\
0   & z   & 0   & 1-z \\
\end{array}
\right)
\]
%--
The transformed invariants are $M^2_i= z\,(1-z)\,u$ and
%--
\bqas
S &=& (1-2\,z)^2\,s, \quad
T = (1-2\,z)^2\,t, \quad
U = u.
\eqas
%--
The solution of $B_4 = 0$ which makes singular the integrand is
%--
\bqas
{}&{}& 2\,u^2\,z_{\ssA\ssT}\,\lpar z_{\ssA\ssT} - 1\rpar =
4\,m^2\,s + u\,t
\nl
&+& \Bigl[ s\,\lpar 4\,m^2-u\rpar\,\lpar 4\,m^2\,s + u\,t\rpar\Bigr]^{1/2}.
\eqas
%--
The effect of the transformation is simple, we have mapped the original box
onto a box which satisfies the condition stated in Coleman - Norton theorem.

As an example of ODE in $z$ we consider the scalar box after the transformation
$P_{1,4}= p_{1,4}+z\,(p_1+p_2)$ and $P_{2,3}= p_{2,3}-z\,(p_1+p_2)$,
%--
\bqa
D^n_0\lpar \{\nu\} \rpar &=& 
\lambda_n\,\int\,d^nq\,
\frac{1}{\prod_{i=0,3}\,[i]^{\nu_i}},
\nl
D^n_0(i) &=& D^n_0\lpar 1,\,\dots\,,2,\,\dots\,,1\rpar 
\nl
D^n_0 &=& D^n_0\lpar 1,\,\dots\,,1\rpar
\nl
\frac{d}{dz}\,D^n_0 &=& 2\,z s\,\Bigl[ D^n_0(2) + D^n_0(4) \Bigr] +
\hbox{triangles}
\eqa
%--
Using IBP-techniques (and dropping the superscript $n$) we derive
%--
\bq
D_0(i) = R^{-1}_{4,ij}\,d_j, \quad \hbox{det}\,R_4(z_{\ssA\ssT}) = 0
\eq
%--
where $d_i$ contains $D_0$ or triangles. Introducing $r= z^2-z$, we obtain
%--
\bq
\frac{d}{dr}\,D_0(r) = C^{-1}_4(r)\,
\Bigl[ X(r)\,D_0(r) + D_{\rm rest}(r) \Bigr]
\eq
%--
where $C_4$ is the corresponding Caley determinant. Furthermore, we have
%--
\bq
\frac{d}{dr}\,C_4 = -\,2\,X(r),
\eq
%--
which leads to the expected solution,
%--
\bq
D_0(r) = 
\frac{D^{\rm sing}}{\lpar r-r_{\ssA\ssT}\rpar^{1/2}} + D^{\rm reg}(r)
\eq
%--
Before turning to a final example it is instructive to consider
the deep connection between ODE for Feynman diagrams, IBP identities and
analytical properties of the diagrams. It can be seen as follows:
for a given set of momenta we consider the transformation
$P_i = T_{ij}(z)\,p_j$, subject to $\sum\,P = \sum\,p = 0$. 
Consider a generalized, scalar, box (arbitrary powers in propagators);
we will also need the IBP equations for $D_0(1,1,1,1)$ and will define
$D_0(1) = D_0(1,2,1,1)$ till $D_0(4)= D_0(2,1,1,1)$. Again, we can use IBP to get
%--
\bq
R_{4\,;\,ij}\,D_0(j) = \delta_{i4}\,D_0\lpar 1,1,1,1\rpar + \Delta\,D_0(i),
\eq
%--
where $\Delta\,D_0(i)$ contains only $3\,$-point functions.
%--
Introduce the Caley determinant $C_4$; it follows that $R_4\,U = 2\,M_4$,
where $U$ is unimodular (a similar relation holds for arbitrary $N$), i.e. 
%--
\bq
{\rm det}\,R_4 = 16\,C_4,
\label{IBPC}
\eq
%--
so that the differential equation for the transformed box is
%--
\bqas
\frac{d}{dz}\,D_0\lpar 1,1,1,1 \rpar &=& \frac{X}{C_4}\,
D_0\lpar 1,1,1,1 \rpar + Y.
\eqas
%--
A straightforward calculation shows that
%--
\bq
X = -\,\frac{1}{2}\,\frac{d}{dz}\,C_4
\eq
%--
for all values of $\{p\},\,\{m\}$ and for an arbitrary transformation $T$.  
\eqn{IBPC} holds for all $N$, i.e. ${\rm det}\,R_{\ssN} = 2^{\ssN}\,C_{\ssN}$, 
where one should remember that in four dimensions $C_{\ssN} = 0$ for $N > 6$. The 
homogeneous term has the general stucture
%--
\bqas
\frac{d}{d z}\,N^n_0 &=& - \frac{1}{2}\,\Bigl[
C^{-1}_{\ssN}\,\frac{d C_{\ssN}}{dz}
\nl
{}&+& \lpar N-n\rpar\,B^{-1}_{\ssN}\,\frac{d B_{\ssN}}{dz}\Bigr]\,
N^n_0 + Y_{\ssN-1},
\eqas
%--
where $Y_{\ssN-1}$ is a combination of $N-1\,$-point integrals.
%--
\section{An explicit example}
%--
Our last case in point is given by the ODE for 
$H \to g(p_1)g(p_2)$ decay amplitude.
Here there is one form factor $F_{\ssD}$ that can be written, without reduction,
as $F_{\ssD} = \sum_i\,F_i$,
%--
\bqa
F_1 &=& \frac{\lambda_n}{2}\,\intmomi{n}{q}\frac{\mhs - 2\,\mts}{[0][1][2]}
\nl
F_2 &=& -\,2\,\lambda_n\,\intmomi{n}{q}\frac{\spro{q}{p_1}}{[0][1][2]}
\nl
(n-2)\,F_3 &=& \lambda_n\,\int\frac{d^n q}{[0][1][2]}\,\Bigl[ (6-n)\,q^2 
\nl
{}&+& \frac{16}{\mhs}\,\spro{q}{p_1}\spro{q}{p_2}\Bigr]
\eqa
%--
Suppose that $\mhs < 4\,\mts$: define the transformation $P_i= T_{ij}\,p_j$
with
%--
\[
T = 
\left(
\begin{array}{cc}
z   & 1-z \\
1-z & z \\
\end{array}
\right)
\]
%--
Then $B \to \mhs\,C/G$ with
$C = r^2 + \mu^2_t\,(1 + 4\,r)$ and $G = -\frac{1}{4}\,\mhs\,(1 + 4\,r)$,
being $r= z\,(z-1)$ and $\mu^2_t\,\mhs = \mts$.
The solution for AT is given by
%--
\bqa
{}&{}& r_{\ssA\ssT} = -\,2\mu^2_t\,\Bigl[ 1 + \sqrt{1-\frac{1}{4\,\mu^2_t}}\;\Bigr]
\nl
{}&{}& -\,\infty < r_{\ssA\ssT} < -\,\frac{1}{2}
\eqa
%--
The corresponding system of ODE will be written in terms of $F_{1,2}$ and 
$F_{\ssD}$
giving
%--
\bq
\frac{d}{d r}\,F_i = X_{ij}\,F_j + Y_j, \quad
i,j = 1,2,D,
\eq
%--
where $X$ and $Y$ are obtained by using IBP techniques, e.g.
%--
\bqa
X_{\ssD 1} &=& X_{\ssD\ssD} = - \frac{2}{1+4\,r}, 
\quad
X_{\ssD 2} = 0,
\eqa
%--
etc, with $Y$ given by generalized two-point functions. Boundary conditions at AT 
are
%--
\bqa
F_{\ssD} &\stackrel{\sim}{\ssA\ssT}& 
\Bigl[ \frac{\mhs}{8}\,\lpar 1 + 6\,r_{\ssA\ssT}\rpar
\nl
{}&-& \mts\,\lpar 1 + 4\,r_{\ssA\ssT}\rpar\Bigr]\,
C^{\rm sing}_0(z),
\nl
F_1 &\stackrel{\sim}{\ssA\ssT}&  \frac{1}{2}\,\lpar \mhs - 2\,\mts \rpar\,
C^{\rm sing}_0(z)  
\nl
F_2 &\stackrel{\sim}{\ssA\ssT}& \mhs\,z_{\ssA\ssT}\,C^{\rm sing}_0(z).
\eqa
%--
More general transformations, not singular for any $z \in R$, exist but will not
discussed here. The pure-scalar term becomes
%--
\bqa
C_0(r) &=& C^{\rm sing}_0(r) + c^{\rm reg}_0(r)
\nl
{}&=& c^{\rm sing}_0(r)\,\ln\frac{B_3(r)}{\mhs} + c^{\rm reg}_0(r)
\nl
\frac{d}{d r}\,c^{\rm sing}_0 &=& -\,\frac{2}{1+4\,r}\,c^{\rm sing}_0
\eqa
%--
with boundary conditions
%--
\bqas
c^{\rm sing}_0\lpar z_{\ssA\ssT}\rpar &=& 
\frac{2\,\pi i}{\mhs}\,\beta\lpar \,r_{\ssA\ssT}\rpar
\;\;\;
\beta^2(r) = 1 - 4\,\frac{\mu^2_t}{r}
\eqas
%--
while the regular part is computed numerically (boundary condition for the regular
part will not be reported here).

The general strategy, e.g. for processes with $N= 4$, is as follows: define
%--
\bqa
D_{n_0\dots n_3}(i) &=& \lambda_n\,\intmomi{n}{q}
\frac{(\spro{q}{q})^{n_0}\,\dots\,(\spro{q}{P_3})^{n_3}}{[0]\,
\dots\,[i]^2\,\dots\,[3]}
\eqa
%--
which satisfy a recurrence relation (IBP)
%--
\bqa
D_{n_0\dots n_3}(i) &=& R^{-1}_{ij}\,d_{n_0\dots n_3}(j) +
d'_{n_0\dots n_3}(i),
\eqa
%--
then find the minimal set of linear combinations $F = c\,D$ such that
${\rm Amp} = \sum\,F$ with $\{F\}$ closed under $d/dz$.
%--
\section{Extension to multi-loop}
%--
Although we shall not discuss higher loops in details here, we present
one simple example: the equal mass two-loop sunset $S$~\cite{Remiddi:2004kv}; with 
scaled masses $\;m= 1$ and $p^2= x\;$ we perform the transformation $\;x \to z\,x$ 
%--
\bqas
{}&{}& x\,z\,\lpar x\,z + 1\rpar\,
\lpar x\,z + 9\rpar\,\frac{d^2}{dz^2}\,S(x,z) =
\nl
{}&{}& P(x,z)\,\frac{d}{dz}\,S(x,z) + Q(x,z)\,S(x,z) + R(x,z)
\eqas
%--
The AT solution is $z_{\ssA\ssT} = -\,x^{-1}\;\;$ (note that here
AT = pseudo-threshold). For different masses we map
%--
\bqa
m_i &\to& M_i = \frac{z - z_{\ssA\ssT}}{1 - z_{\ssA\ssT}}\,m_i +
\frac{1 - z}{1 - z_{\ssA\ssT}}\,m,
\eqa
%--
and use the previous calculation of AT.
%--
\section{Conclusions}
In conclusions we have presented a proposal for solving the problem
of reducing Feynman diagrams which is based on a single variable deformation 
of the amplitude.
%--

%===
\end{document}